\documentclass{webofc}
\usepackage[varg]{txfonts}   
\newcommand{\pt}           {\ensuremath{p_{\rm T}}\xspace}

\newcommand{\RAA}         {\ensuremath{R_{\rm AA}}\xspace}

\newcommand{\nineH}        {$\sqrt{s}~=~0.9$~Te\kern-.1emV\xspace}
\newcommand{\seven}        {$\sqrt{s}~=~7$~Te\kern-.1emV\xspace}
\newcommand{\twoH}         {$\sqrt{s}~=~0.2$~Te\kern-.1emV\xspace}
\newcommand{\twosevensix}  {$\sqrt{s}~=~2.76$~Te\kern-.1emV\xspace}
\newcommand{\five}         {$\sqrt{s}~=~5.02$~Te\kern-.1emV\xspace}
\newcommand{\thirteen}     {$\sqrt{s}~=~13$~Te\kern-.1emV\xspace}
\newcommand{\twosevensixnn}{$\sqrt{s_{\mathrm{NN}}}~=~2.76$~Te\kern-.1emV\xspace}
\newcommand{\fivenn}       {$\sqrt{s_{\mathrm{NN}}}=5.02$~Te\kern-.1emV\xspace}

\newcommand{\GeVc}         {Ge\kern-.1emV/$c$\xspace}
\newcommand{\MeVc}         {Me\kern-.1emV/$c$\xspace}
\newcommand{\TeV}          {Te\kern-.1emV\xspace}
\newcommand{\GeV}          {Ge\kern-.1emV\xspace}
\newcommand{\MeV}          {Me\kern-.1emV\xspace}
\newcommand{\GeVmass}      {Ge\kern-.2emV/$c^2$\xspace}
\newcommand{\MeVmass}      {Me\kern-.2emV/$c^2$\xspace}

\newcommand{\jpsi}         {\ensuremath{\mathrm{J}/\psi}\xspace}
\newcommand{\psitwos}      {\ensuremath{\psi {\rm (2S)}}\xspace}

\newcommand{\posyint}{\ensuremath{ 2.5 < y < 4}~}

\begin{document}

\title{Coherent $\jpsi$ photoproduction and polarization in peripheral Pb$\--$Pb collisions}
\author{\firstname{Afnan} \lastname{Shatat}\inst{1}\fnsep\thanks{\email{afnan.shatat@cern.ch}} 
for the ALICE Collaboration}
\institute{Universit\'e Paris-Saclay, CNRS/IN2P3, IJCLab, 91405 Orsay, France}

\abstract{Photonuclear reactions are induced by the strong electromagnetic field generated by ultrarelativistic heavy-ion collisions.
These processes have been extensively studied in ultraperipheral collisions, where the impact parameter is larger than twice the nuclear radius.
In recent years, the observation of coherent $\jpsi$ photoproduction at very low transverse momentum has been claimed in nucleus$\--$nucleus (A$\--$A) collisions with nuclear overlap, based on the measurement of an excess in the $\jpsi$ yield with respect to hadroproduction expectations.
Such quarkonium measurements can help to constrain the nuclear gluon distribution at low Bjorken-$x$ and high energy. The photoproduced quarkonium is expected to preserve the polarization of the incoming photon due to the $s$-channel helicity conservation. 
In this contribution, we report on the new preliminary measurements of the $y$-differential cross section and the first polarization analysis of coherently photoproduced $\jpsi$ in peripheral Pb$\--$Pb collisions with the ALICE detector at forward rapidity in the dimuon decay channel. Comparison with models will be shown when available. 
}
\maketitle
\section{Introduction}
\label{sec:intro}
A coherent $\jpsi$ photoproduction observation has been reported for the first time by the ALICE Collaboration in nucleus$\--$nucleus (A$\--$A) collisions with nuclear overlap at very low transverse momentum ($\pt$)~\cite{ALICE:2015mzu}.
Coherent vector meson photoproduction is a well-established process in ultraperipheral collisions (UPCs),
but its observation in Pb$\--$Pb collisions with nuclear overlap raises several theoretical challenges, for instance on the survival of the coherence condition when the colliding nuclei are broken by the hadronic interaction. 
This process is also a valuable tool for studying the nuclear gluon distribution in the poorly-known region of low Bjorken-$x$~\cite{Eskola:2009uj}.
The determination of $\jpsi$ polarization at low $\pt$ is crucial in verifying the origin of the observed excess yield. Due to the conservation of s-channel helicity, the produced quarkonium is expected to preserve the polarization of the incoming photon~\cite{GILMAN1970387}.
The coherent $\jpsi$ photoproduction cross section was measured as a function of centrality in Pb$\--$Pb collisions with nuclear overlap~\cite{ref:Photo_cent_5}.
Reproducing the collision-centrality dependence of the $\jpsi$ photoproduction cross section remains challenging for most of the available models, as the photon flux and the photonuclear cross sections used as input to the calculations can strongly be affected by the presence of a nuclear overlap. To further constrain theoretical models, additional differential measurements are required. For instance, a strong rapidity ($y$) dependence is expected from models (see for example~\cite{ref:Model_GBWIIM}), especially in the forward rapidity region $2.5 < y <4$.
\section{Experiment and analysis}
The ALICE detector is described in detail in ~\cite{ALICE:2008ngc}.
In the presented results, the $\jpsi$ is studied in the dimuon decay channel with the ALICE muon spectrometer.
The muon spectrometer consists of five tracking and two triggering stations. An iron wall is located between the last tracking and the first triggering station to filter muons.
A front absorber is located between the interaction point and the first tracking station, to suppress the hadronic background. 
The cross section and polarization analyses are conducted in Pb$\--$Pb collisions at \fivenn in the  70$\--$90$\%$ centrality range. The event and track selections are the same as in the previous measurement~\cite{ref:Photo_cent_5}. The used data sample has an integrated luminosity of $756~\pm~19~\mu\rm{b}^{-1}$.
The $\jpsi$ yield is extracted by fitting the opposite sign dimuon invariant mass distribution. The polarization is measured through a fit to the dimuon angular distribution in order to extract the three polarization parameters ($\lambda_\theta, \lambda_\phi, \lambda_{\theta \phi}$) in the helicity frame where the $\jpsi$ flight direction in the lab frame is the polarization axis. The set of parameters  ($\lambda_\theta, \lambda_\phi, \lambda_{\theta \phi}$) = (1, 0, 0) corresponds to transverse polarization.
A further explanation of the polarization parameters and the fit function is available in~\cite{ref:UPC_transvPol}.

\section{Results and discussion} 
\label{sec:results}
\begin{figure*}[!hbt]
\centering
\includegraphics[width = 0.49\textwidth]{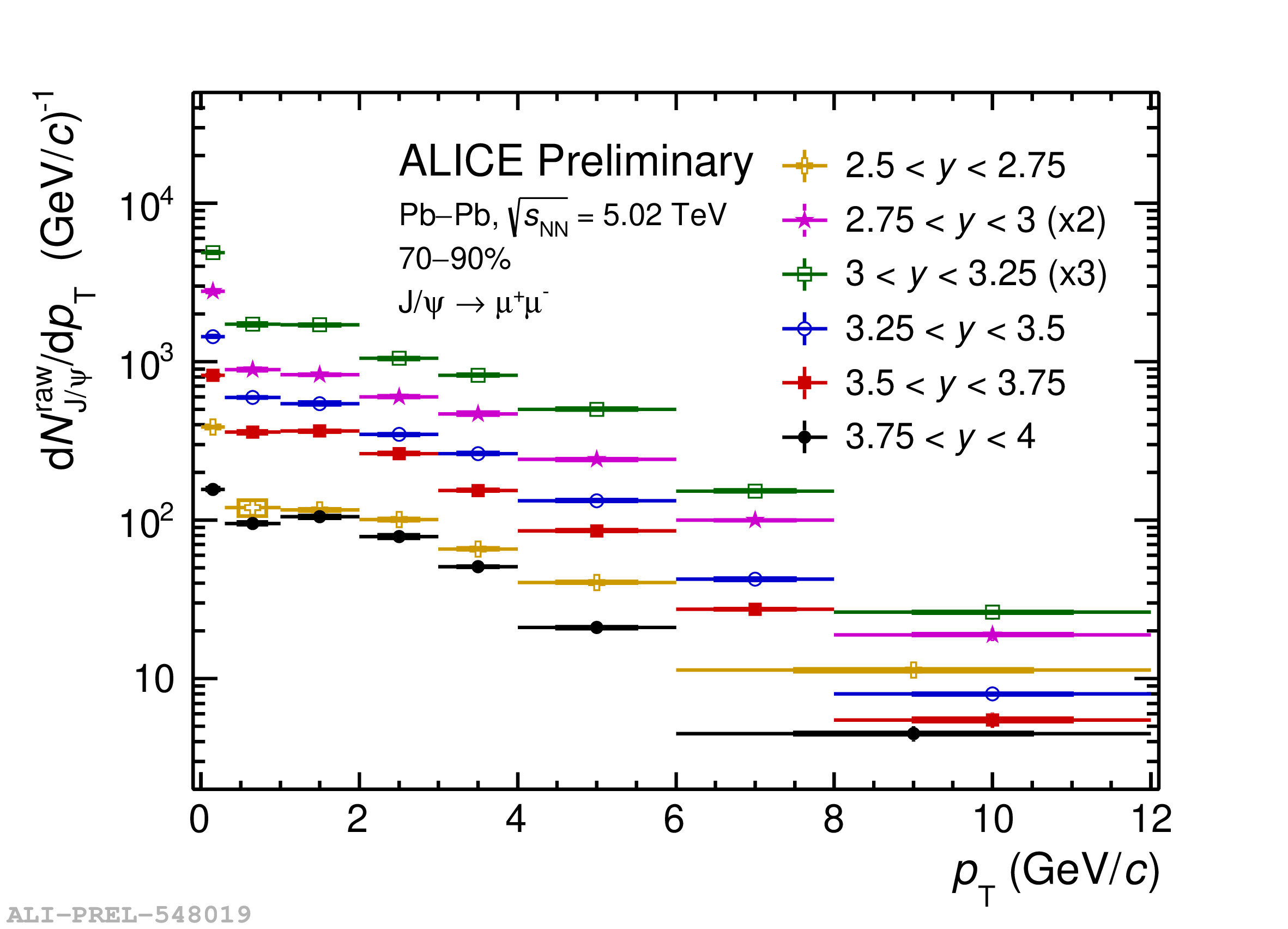}
\includegraphics[width = 0.49\textwidth, height =0.35\textwidth]{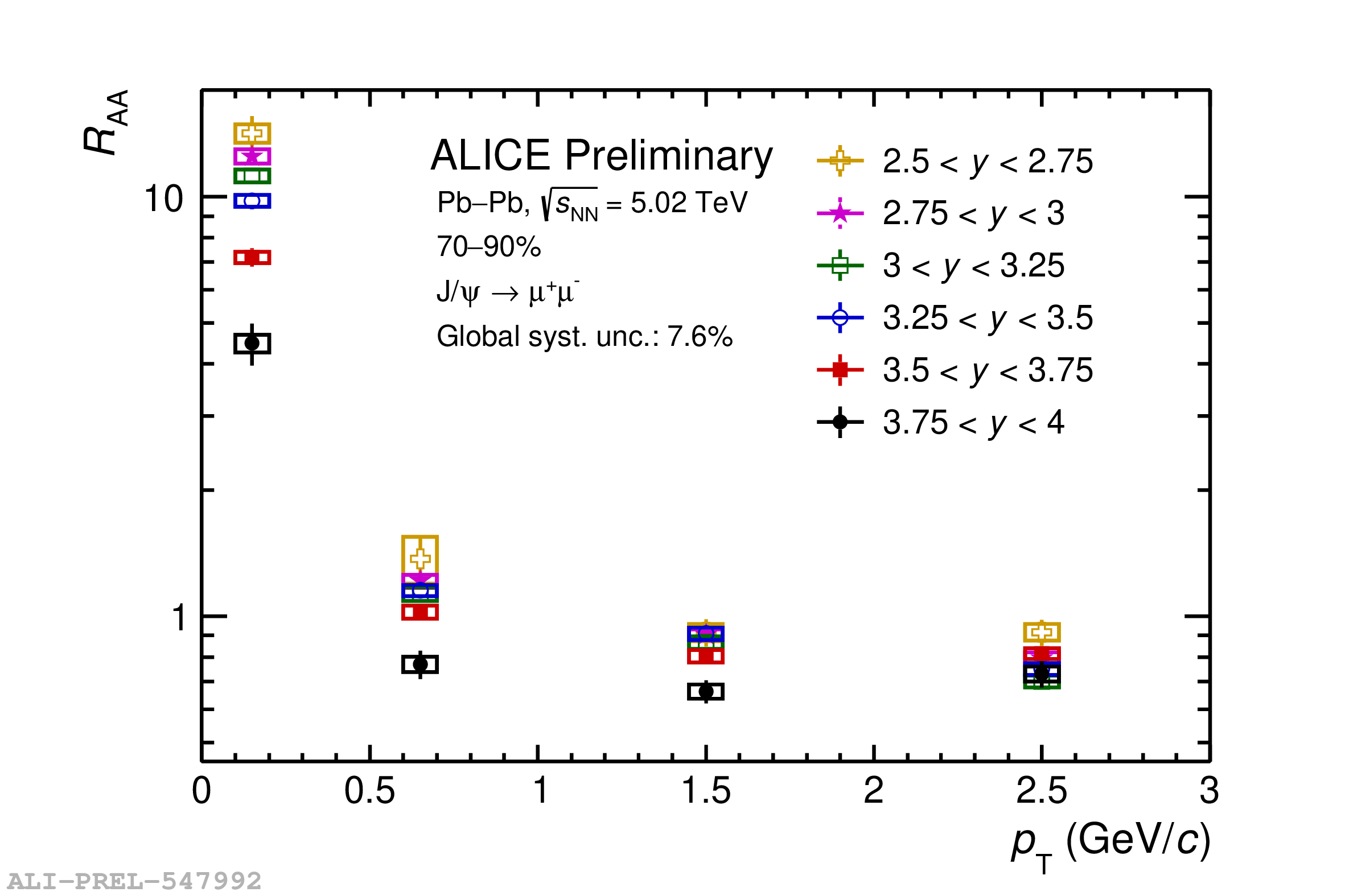}
\caption{The $\pt$-differential $\jpsi$ raw yield distribution (left) and the $\jpsi$ nuclear modification factor $\RAA$ as a function of $\pt$ (right) in Pb$\--$Pb collisions at \fivenn in the 70$\--$90$\%$ centrality range, for several rapidity intervals.}
\label{fig:Signalinfo} 
\end{figure*}

The $\jpsi$ raw yield is extracted in six rapidity intervals at forward rapidity $\posyint$ in Pb$\--$Pb collisions at \fivenn in the 70$\--$90$\%$ centrality range. The resulting $\pt$-differential $\jpsi$ raw yield is displayed in the left panel of Fig.~\ref{fig:Signalinfo}.
The right panel of Fig.~\ref{fig:Signalinfo} shows the measurement of the corresponding nuclear modification factor \mbox{$\RAA^{\jpsi} = (1/ <T_{\rm AA}>) \times dY_{\rm{AA}}^{\jpsi}/d{\pt}/d\sigma_{\rm{pp}}^{\jpsi}/d{\pt}$}, where $<T_{\rm AA}>$ and $Y_{\rm{AA}}^{\jpsi}$ are the average nuclear overlap function and the invariant $\jpsi$ yield in A$\--$A collisions, respectively, while $\sigma_{\rm{pp}}^{\jpsi}$ is the hadronic $\jpsi$ cross section in proton$\--$proton collisions at the same energy.
The $\RAA$ decreases as the rapidity increases and exhibits a significant rise at low $\pt$.

For $\pt < 0.3  $~\GeVc, the $\jpsi$ hadronic yield in Pb$\--$Pb collisions is estimated via a data driven approach that uses as inputs the $\jpsi$ hadronic cross section in pp collisions, the $\jpsi~\RAA$, and the hadronic $\jpsi$ Acceptance x Efficiency ($A\times\epsilon$) in Pb$\--$Pb collisions. A significant $\jpsi$ excess over the expected hadronic yield is observed in the raw $\jpsi$ yield distribution for $\pt < 0.3  $~\GeVc. The excess yield along with the rise in the $\RAA$ indicates that photoproduction occurs at low $\pt$.
To obtain the coherent $\jpsi$ photoproduction yield, the excess yield is then corrected for the fraction of incoherent $\jpsi$ ($f_I$) and  coherent $\jpsi$ from coherent $\psitwos$ feed down ($f_D$), taken from UPC measurements~\cite{ref:UPC_fIfDvalues}.
In order to calculate the coherent $\jpsi$ photoproduction cross section, the coherent yield is corrected for $A\times\epsilon$ assuming transverse polarization and divided by the $\jpsi$ branching ratio in the dimuon decay channel and the integrated luminosity.

\begin{figure*}[!bt]
\centering
\includegraphics[width = 0.49\textwidth]{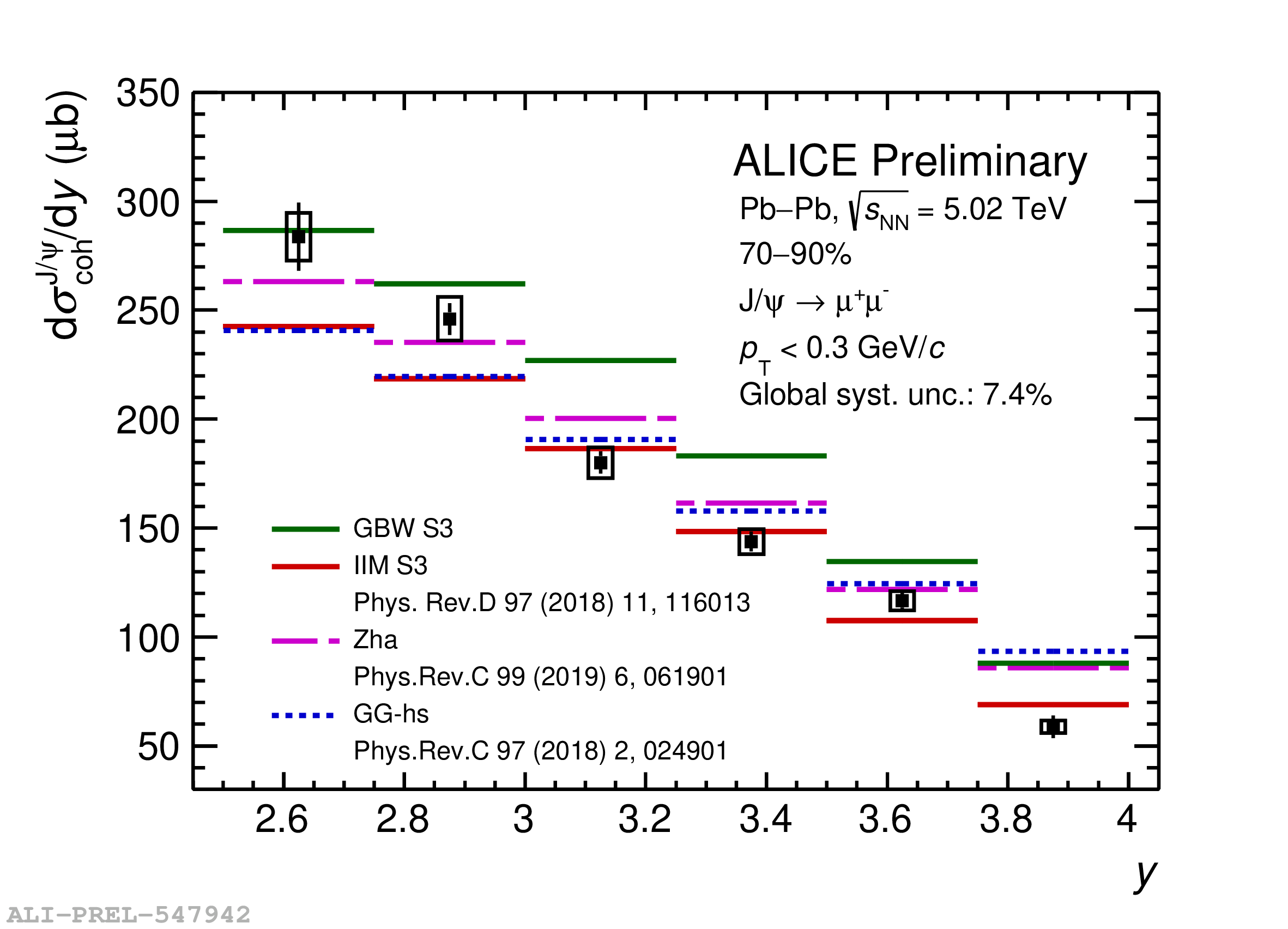}
\includegraphics[width = 0.49\textwidth]{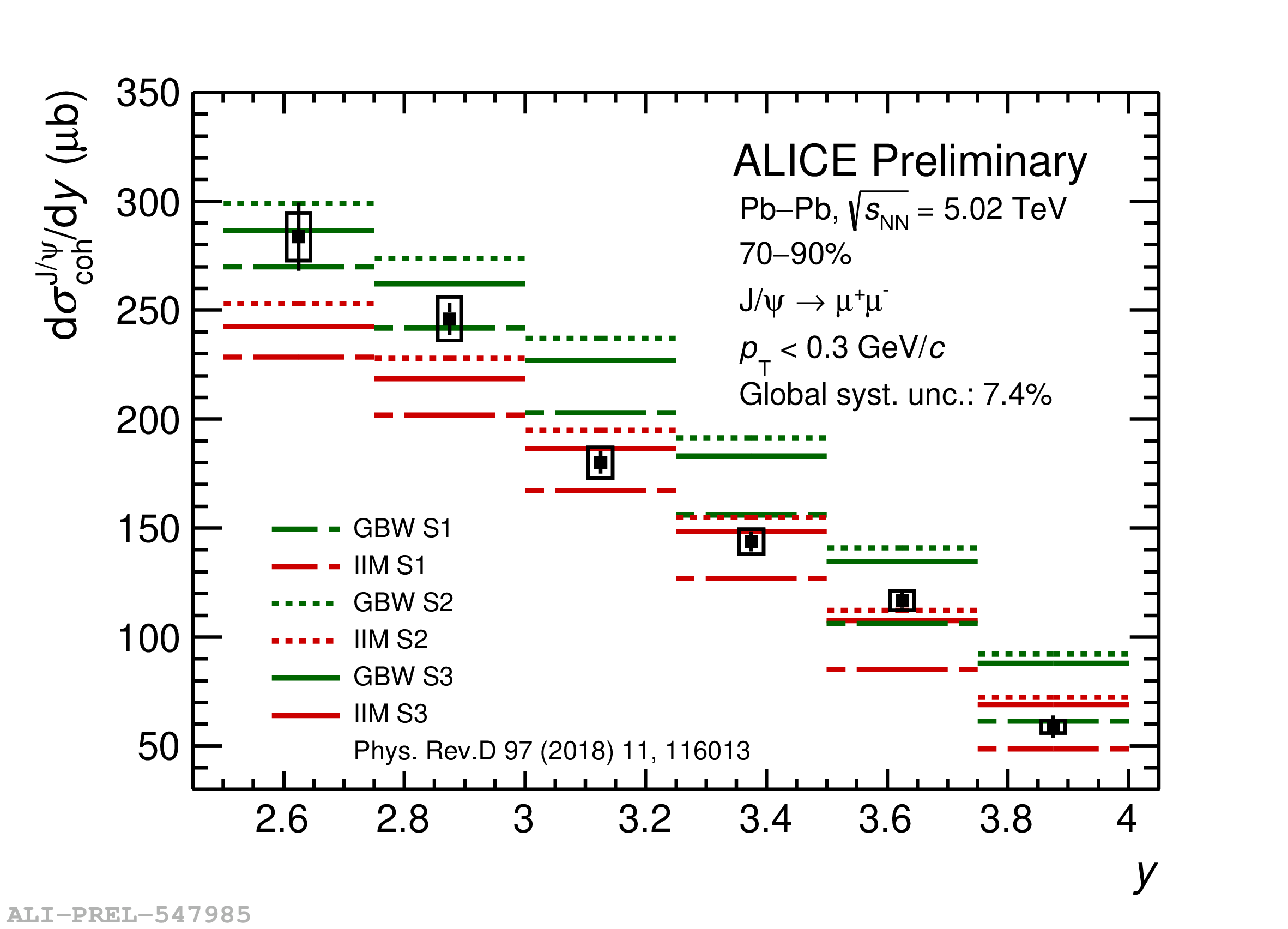}
\caption{The forward $y$-differential coherent $\jpsi$ photoproduction cross section in Pb$\--$Pb collisions at \fivenn in the 70$\--$90$\%$ centrality range, and for $\pt < 0.3  $~\GeVc, compared to GG-hs~\cite{ref:Model_GGhs}, Zha~\cite{ref:Model_Zha}, and GBW/IIM~\cite{ref:Model_GBWIIM} (left) and the three scenarios of the GBW/IIM~\cite{ref:Model_GBWIIM} model (right).}
\label{fig:crosssection} 
\end{figure*}

Several $\jpsi$ photoproduction models in UPC were extended to describe peripheral collisions. First, in the hot-spot model (GG-hs), the photon flux is estimated in the same way as in the UPC case, but the integral is limited to the impact parameter range of the selected centrality class~\cite{ref:Model_GGhs}.
Second, based on a picture of photon-pomeron coupling, the Zha model proposes that one nucleus acts as a photon emitter and the spectator nucleons in the other nucleus, which do not undergo the hadronic interaction, act as a pomeron emitter~\cite{ref:Model_Zha}.
Finally, the GBW/IIM model considers an extension from the UPC calculations to peripheral collisions with three distinct scenarios~\cite{ref:Model_GBWIIM}.
The GBW and IIM stand for two different calculations for the photon-target interaction within a dipole picture framework.
Scenario 1 (S1) has no relevant modification in relation to the ultraperipheral regime. 
In S2, an effective photon flux is considered including only the photons that reach the spectator region of the target nucleus. In S3, the effective photon flux of S2 is used, and in order to exclude the nuclear overlap region, an extra modification of the UPC photonuclear cross section is considered. 
In Fig.~\ref{fig:crosssection}, the $y$-differential coherent $\jpsi$ photoproduction cross section is compared to several models.
In peripheral collisions, where the nuclear overlap effect on the coherent $\jpsi$ photoproduction cross section is expected to be small, some sets of models which include (Zha, IIM S2/S3) or not (GG-hs, GBW S1) the nuclear overlap can both describe quantitively enough the overall data trend (magnitude and rapidity dependence of the cross section). The current theoretical calculations would need further constrain from UPC results~\cite{ref:UPC_sigma_y} in the first place, to better interpret the peripheral data.

\begin{figure*}[!hbt]
\centering
\includegraphics[width = 0.4\textwidth]{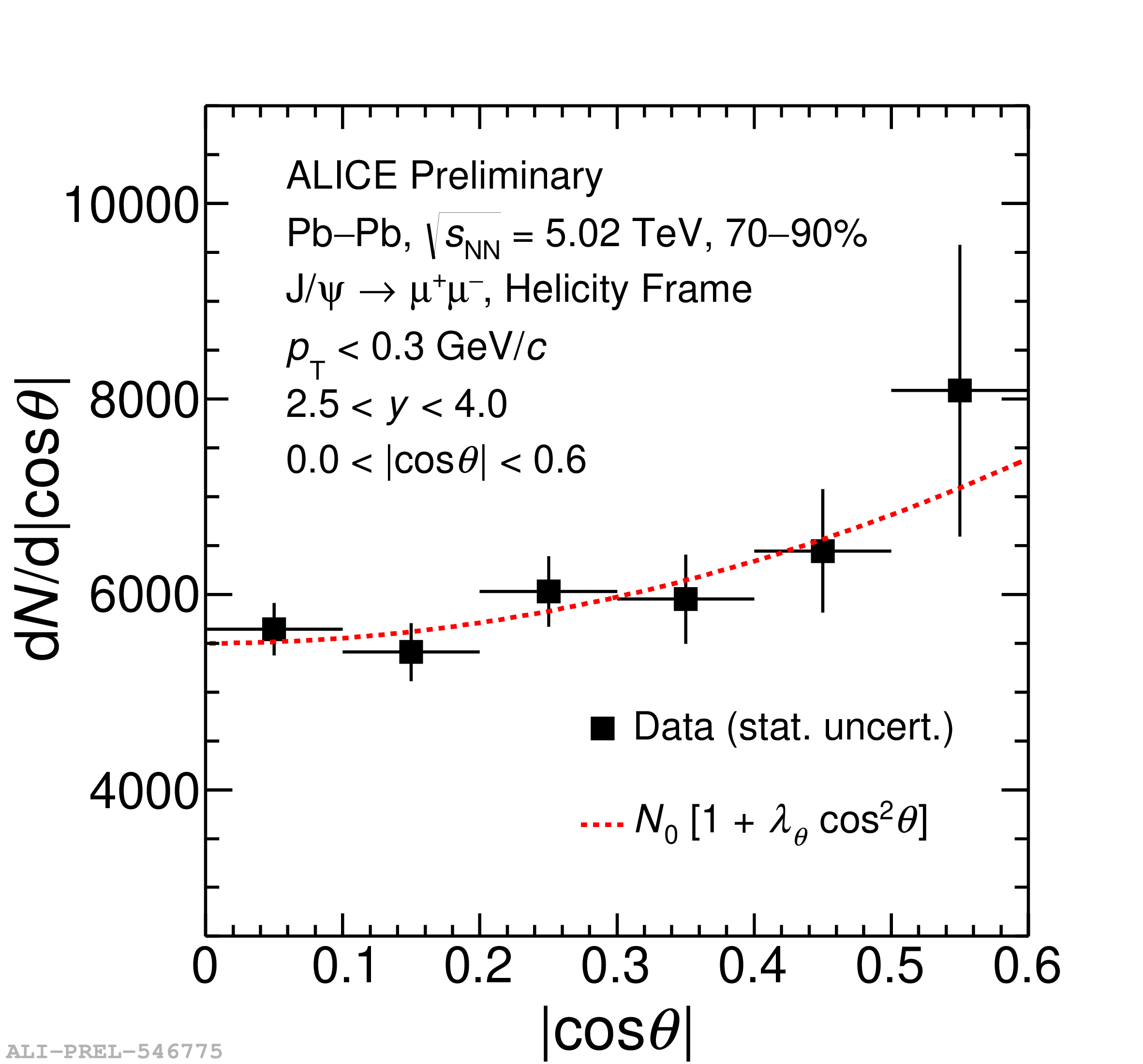}
\includegraphics[width = 0.4\textwidth]{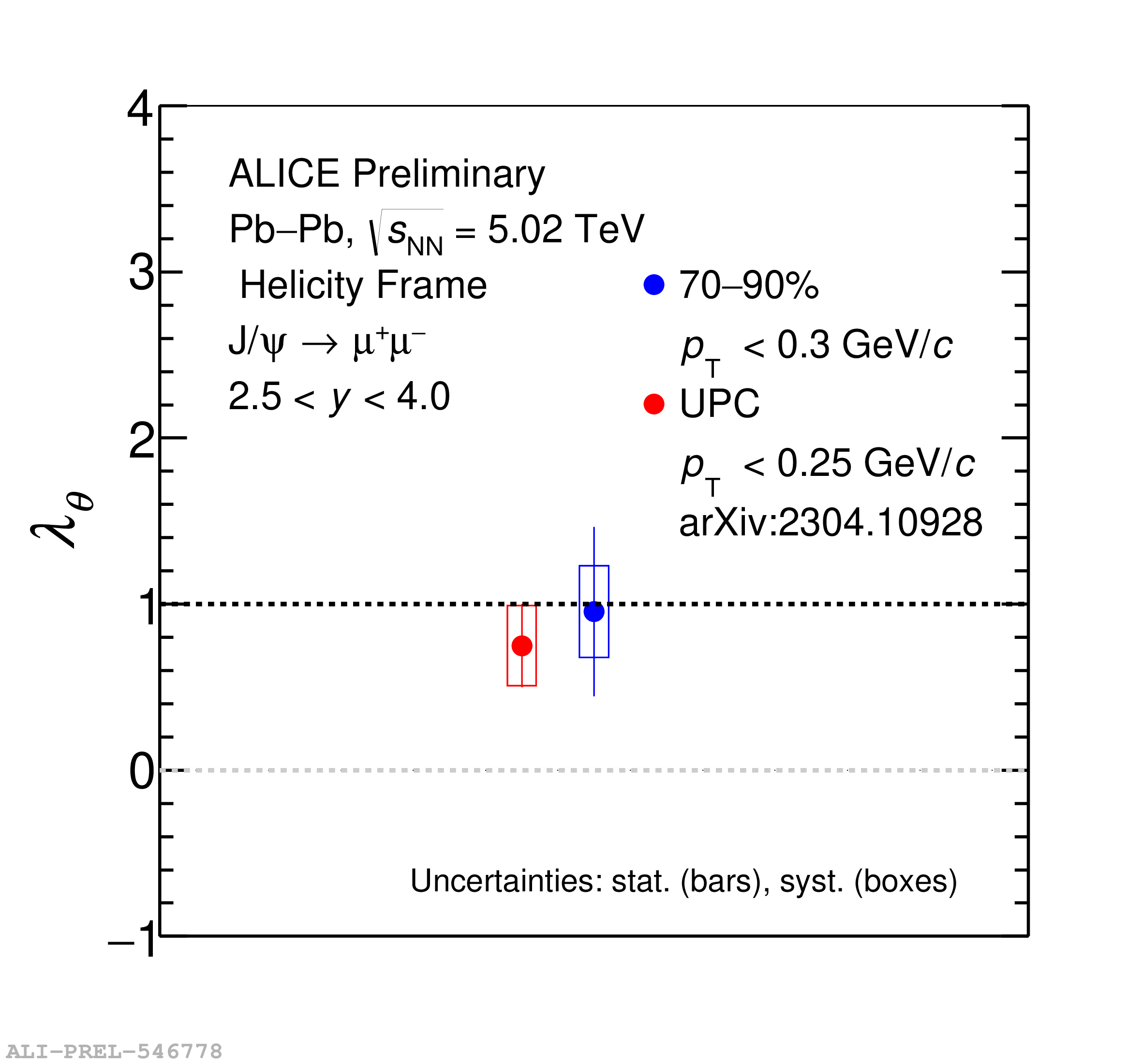}
\caption{Inclusive $\jpsi$ cos$\theta$ distribution in the helicity frame for $\pt<0.3$~\GeVc  and  $\posyint$  in Pb$\--$Pb collisions at \fivenn (left). The polarization parameter $\lambda_\theta$ extracted from the left plot is compared to the coherently photoproduced $\jpsi$ $\lambda_\theta$ polarization parameter in UPC~\cite{ref:UPC_transvPol} (right).}
\label{fig:polarization} 
\end{figure*}

The dimuon angular distribution in cos$\theta$ intervals and for $\pt < 0.3$~\GeVc is presented in the left panel of Fig.~\ref{fig:polarization}.
The polarization parameter $\lambda_\theta$ is extracted from a fit to the angular distribution and is compared to the UPC measurement in the right panel of Fig.~\ref{fig:polarization}.
The parameter $\lambda_\theta$ is close to unity and is compatible with the observed polarization for the coherently photoproduced $\jpsi$ in UPC within the large uncertainties~\cite{ref:UPC_transvPol}. 
This observation supports the scenario of a dominant coherent $\jpsi$ photoproduction contribution over the hadronic one, in the kinematic region $\pt < 0.3$~\GeVc and  $\posyint$, in peripheral Pb$\--$Pb collisions~\cite{ref:Photo_cent_5}. 
\section{Conclusion} 
\label{sec:conclusion}
 We have reported the first $y$-differential coherent $\jpsi$ photoproduction cross section measurement along with the first inclusive $\jpsi$ polarization measurement, for $\pt < 0.3$~\GeVc and forward rapidity, in peripheral Pb$\--$Pb collisions at \fivenn with ALICE. 
The cross section measurement was compared to several models that can reproduce the magnitude of the cross section, but fail at reproducing the $y$-dependence,  similarly to the UPC case.
The existence of coherent $\jpsi$ photoproduction in Pb$\--$Pb collisions with nuclear overlap therefore remains a theoretical challenge. Extending the cross section measurement to the most central collisions can serve as a discriminator of the proposed models.
The polarization parameter $\lambda_\theta$, measured in the helicity frame, is compatible with transversely polarized $\jpsi$ and is consistent with the measurement of the coherently photoproduced $\jpsi$ in UPCs.
A larger Pb$\--$Pb data sample of about $6~\rm{nb}^{-1}$ integrated luminosity is expected to be collected during LHC Run 3.
This will allow one to obtain a significant signal of coherently-photoproduced $\jpsi$ in the most central collisions and to improve the precision of the cross section and polarization measurements. In addition, it will permit the examination of other coherently-photoproduced vector mesons to investigate potential QGP effects on these photoproduced probes. 
\bibliography{bibliography}
\end{document}